\documentstyle[aps,pra,eqsecnum,amssymb]{revtex}
\begin{document}
\draft
\title{Separable approximation for mixed states of composite quantum
systems}
\author{Thomas Wellens$^{1,2}$, Marek Ku\'s$^1$}
\address{$^1$Center for Theoretical Physics, Polish Academy of Sciences,
al. Lotnik\'{o}w 32/46, PL-02-668 Warszawa, Poland, \\
$^2$MPI f\"ur Physik komplexer Systeme, N\"othnitzer Str.\ 38, D-01187
Dresden, Germany}

\date{April 23, 2001}

\maketitle

\begin{abstract}
We describe a purely algebraic method for finding the best separable
approximation \cite{lewenstein} to a mixed state of a
composite $2\times 2$
quantum system, consisting of a decomposition of the state into a linear
combination of a mixed separable part and a pure entangled one.
We prove that, in a generic case, the weight of the pure part in the
decomposition equals the concurrence of the state.
\end{abstract}

\pacs{03.67.-a,03.65.Ta,89.70.+c}

\section{Introduction}

One of the main interests of quantum information theory concerns the
nonclassical features connected with the nonseparability (or entanglement)
of states of composite systems. Since entanglement plays a crucial role in
various
applications in quantum information processing, the problem of characterization
of entangled states is of a paramount importance.

It is easy to check whether a {\em pure} state of a composite system is
separable or
entangled. The situation complicates in the case of mixed states. A simple and
practical necessary criterium of separability is known, but there are no known
sufficient conditions for higher dimensional composite systems.

Recently \cite{lewenstein}, a very interesting description of entanglement was
achieved by defining the best separable approximation (BSA) of a mixed
entangled state. In the simplest case of a $2\times 2$ dimensional composite
system,
it consists of a decomposition of the state into a linear combination
of a mixed separable part and a pure entangled one. In this way, the whole
nonseparability properties are concentrated in the pure part. It also provides
a natural measure of entanglement given by the entanglement of the pure
part (well defined for the pure states) multiplied by the weight of the pure
part in the decomposition.

In the original paper \cite{lewenstein}, the authors proposed a numerical
method
for finding the BSA in $2\times 2$ systems. Some analytical results
for special
states were found in \cite{englert}. In this paper, we show how to find the BSA
of an arbitrary $2\times 2$ state $\rho$ in a purely algebraic way, without
employing any maximization or optimization
procedure. As a byproduct, we prove that,
in the case that the BSA $\rho_s$ of $\rho$ is of rank $4$, the weight with
which the entangled part enters the decomposition leading to the BSA, equals
another measure of entanglement, namely the concurrence of $\rho$ 
\cite{wootters}. Furthermore, the pure part is maximally entangled in this
case (the last fact was recently proved by other means in \cite{karnas}).

The situation is more complicated if the BSA $\rho_s$ is not of full rank.
As we will show, for $\text{rank}(\rho)=4$ but
$\text{rank}(\rho_s)<4$ the components
of the BSA are determined by a set of two nonlinear equations which can be
easily solved numerically, whereas the case of a degenerate $\rho$
(i.e. $\text{rank}(\rho)<4$)
can be treated as a limiting case of the full-rank one.
It is to stress that in these cases there is
no simple relation between the concurrence of the state and
the weight of the entangled part as we were able to prove for
$\text{rank}(\rho_s)=4$. Presently, we do not have a simple explanation or
interpretation of this fact which deserves further investigations.

The paper is organized as follows. In Section~\ref{sec:sep} we give a maximally
shortened account on separability and entanglement of mixed states. The main
results of the paper are formulated and proved in Section~\ref{sec:bsa}. The
technical lemmas used in the proofs of the 
two main theorems of Section~\ref{sec:bsa}
are relegated to two Appendices - the first contains some more general theorems
concerning properties of mixed states of $2\times 2$ systems, whereas the
second one is mainly devoted to a technical lemma concerning relations
between spectra of two important matrices obtained from the initial mixed
state.

\section{Separability and entanglement of mixed states}
\label{sec:sep}
A mixed state $\rho$ of a bipartite quantum system is separable if it is a
convex combination of product states \cite{werner}
\begin{equation}
\rho=\sum_i^kp_i\rho_i^A\otimes\rho_i^B,\quad p_i\geq 0, \quad
\sum_i^kp_i=1,
\end{equation}
where $\rho_i^A$, $\rho_i^B$ are legitimate (i.e. hermitian and positive
definite) density matrices of the subsystems.

As observed in \cite{peres}, a necessary condition for separability
of $\rho$ is that its {\em partial transposition}, defined as 
\begin{equation}
\rho^{T_B}:=\sum_i^kp_i\rho_i^A\otimes(\rho_i^B)^{T},
\label{ptb}
\end{equation}
is positive definite, i.e.\ is also a legitimate density matrix for the
composite system. (Here, we define the
operation of partial transposition by Eq~(\ref{ptb}) also in the case of an
arbitrary, not necessarily separable state, when $\rho_i^A$ and $\rho_i^B$ do 
not
need to be positive or/and $p_i$ are not all positive - such a decomposition
obviously exists for an arbitrary $\rho$).
For low dimensional ($2\times 2$ and $2\times 3$) systems
the above condition is also sufficient \cite{horodecki3}.

Obviously, the result of partial transposition depends on the 
basis in subspace ${\mathcal H}_B$.
If we change the bases of ${\mathcal H}_A$ and ${\mathcal
H}_B$ by a local transformation
$U\otimes V$, i.e.\
by unitary rotations $U$ and $V$ in the spaces ${\mathcal H}_A$ and ${\mathcal
H}_B$
respectively (in fact, since the overall phase factor does
not play any role, we can assume $\text{det}U=1=\text{det}V$, i.e.
$U,V\in SU(2)$), the matrix $\rho$ will be transformed according to
\begin{equation}
\rho^\prime=U\otimes V\rho (U\otimes V)^\dagger=
\sum_i^kp_iU\rho_i^A U^\dagger\otimes V\rho_i^B V^\dagger.\label{transf2}
\end{equation}
Consequently, the partial transposition gives
\begin{equation}
\rho^{\prime T_B}=\sum_i^kp_iU\rho_i^A U^\dagger\otimes(V\rho_i^B
V^\dagger)^T=
U\otimes V^*\rho^{T_B}(U\otimes V^*)^\dagger,
\label{transf}
\end{equation}
where asterix denotes the complex conjugation. From (\ref{transf}) it
follows
that the spectrum of $\rho^{T_B}$ is basis-independent.

Observe also the following form of the definition of partial transpose
\begin{equation}
\langle e,f|\rho|e, f\rangle=\langle e,f^*|\rho^{T_B}|e, f^*\rangle,
\label{efref}
\end{equation}
where $|e, f\rangle$ denotes the product vector $|e\rangle\otimes|f\rangle$.

In order to quantify the degree of entanglement of two qubit systems, the
{\em concurrence} was introduced in \cite{wootters}, defined as
\begin{equation}
c(\rho)={\rm max}\{0,c_1-c_2-c_3-c_4\},
\label{conc}
\end{equation}
where $c_1\geq c_2\geq c_3\geq c_4$ are the square roots
of the (real and positive) eigenvalues of the matrix
\begin{equation}
X:=\Sigma\rho^*\Sigma\rho,\quad
\Sigma=\sigma_2\otimes\sigma_2,\quad \sigma_2:=
\left[
\begin{array}{cc}
0 & -i \\
i & 0
\end{array}
\right]
\label{conc2}
\end{equation}

It is a matter of a straightforward calculation to prove that the
concurrence
of a pure state,
\[
|\psi\rangle=a_1|00\rangle+a_2|01\rangle+a_3|10\rangle+a_4|11\rangle=
[a_1,a_2,a_3,a_4]^T,
\]
equals
\begin{equation}
c(\psi)=2|a_1a_4-a_2a_3|=|\langle\psi|\Sigma|\psi^*\rangle|.
\label{purec}
\end{equation}
Due to the normalization condition
$1=\langle\psi|\psi\rangle=|a_1|^2+|a_2|^2+|a_3|^2+|a_4|^2$, we have $0\leq
c(\psi)\leq 1$. The maximum $c(\psi)=1$ is attained for the states called
maximally entangled. The degree of entanglement (i.e.\ the concurrence) is
invariant with respect to local unitary transformations (i.e.\
transformations
of the form $U\otimes V$).

By local transformation, a pure state can be brought to its Schmidt form
$|\psi\rangle=\lambda_1e_1\otimes f_1+\lambda_2 e_2\otimes f_2$, where
$\{e_1,e_2\}$ and $\{f_1,f_2\}$ are appropriately chosen bases in ${\mathcal
H}_A$ and ${\mathcal H}_B$. In these bases thus
$\psi=[\lambda_1,0,0,\lambda_2]^T$ and it is
easy to show that the most general form of a maximally entangled state in
the
original bases reads
\begin{equation}
|\psi\rangle=a_1|00\rangle+a_2|01\rangle\mp (a_2^*|10\rangle -
a_1^*|11\rangle)=
\left[
\begin{array}{c}
a_1\\
a_2\\
\mp a_2^*\\
\pm a_1^*
\end{array}
\right], \quad |a_1|^2+|a_2|^2=\frac{1}{2}.
\label{megen}
\end{equation}

\section{Best separable approximation}
\label{sec:bsa}

Let $\rho$ be a generic density matrix for a two qubit system, i.e. a strictly
positive definite (i.e. rank 4) $4\times 4$ hermitian matrix of unit trace.
According to \cite{lewenstein}, $\rho$ has a unique decomposition of the
form:
\begin{equation}
\rho=(1-\lambda)\left|\psi\right\rangle\left\langle\psi\right| +
\lambda\rho_s,
\label{lewsan}
\end{equation}
where $\rho_s$ is a separable density matrix, $\left|\psi\right\rangle$ is a
pure entangled state, and the parameter $\lambda\in[0,1]$ is maximal.
In the following, we will refer to Eq. (\ref{lewsan}) as the
{\em optimal decomposition} of $\rho$. The separable part
$\rho_s$ is called the
{\em best separable approximation} (BSA) of $\rho$, and $\lambda$ its
{\em separability}.
In this section, we will prove the following

\noindent {\bf Theorem 1. }
Let $\rho$ be an entangled state with ${\rm rank}(\rho)=4$. 
$\rho=(1-\lambda)\left|\psi\right\rangle\left\langle\psi\right| + 
\lambda\rho_s$ is the optimal decomposition of $\rho$ if and only if:

${\rm rank}(\rho_s^{T_B})=3$, i.e. 
$\exists_{|\phi\rangle}~\rho_s^{T_B} |\phi\rangle=0$, and either
\begin{itemize}
\item[(i)]$\exists_{\alpha>0}~[|\phi\rangle\langle\phi|]^{T_B}\
|\psi\rangle=-\alpha
|\psi\rangle$, or
\item[(ii)]${\rm rank}(\rho_s)=3$, i.e. $\exists_{|\tilde{\phi}\rangle}~\rho_s
|\tilde{\phi}\rangle=0$,
and $\exists_{\alpha,\nu\geq 0}~
\left[\nu |\tilde{\phi}\rangle\langle\tilde{\phi}|+
\left[|\phi\rangle\langle\phi|\right]^{T_B}\right]\
|\psi\rangle=-\alpha |\psi\rangle$.
\end{itemize}
According to Lemma 2 of Appendix A, $|\psi\rangle$ is maximally entangled
in case (i).

The first condition, ${\rm rank}(\rho_s^{T_B})=3$ simply states that the
BSA $\rho_s$ lies on the
boundary between the set of separable and the set of entangled states, whereas 
conditions (i) and (ii) describe the relation between the entangled and
separable part of the optimal decomposition. Remarkably, the only 
relevant properties of $\rho_s$ are
the vectors $|\phi\rangle$, and possibly $|\tilde{\phi}\rangle$, in the
kernels of $\rho_s^{T_B}$ and $\rho_s$.
 
Theorem 1 allows us to check
immediately if a given decomposition of $\rho$ is the optimal one.
It also simplifies the construction of the BSA for a given
$\rho$. Indeed, case (i), i.e. any BSA with rank 4, can be solved explicitly,
according to the following

\noindent {\bf Theorem 2. }
If the BSA $\rho_s$
of $\rho$ has full rank, the vector $|\phi\rangle$ in the 
(one-dimensional, see Theorem 1) kernel of
$\rho_s^{T_B}$ is an
eigenvector of 
\begin{equation}
Y=\Sigma\rho^{T_A}\Sigma\rho^{T_B}\label{y}
\end{equation}
belonging to the smallest eigenvalue $\gamma$ of $Y$.
The weight $1-\lambda$ of the entangled part in the optimal decomposition 
is given by $1-\lambda=2\sqrt{\gamma}=c(\rho)$, where $c(\rho)$ is 
the concurrence of $\rho$.

Theorem 2 provides a connection between the BSA and the concurrence of $\rho$,
which was originally \cite{hill} introduced as an auxiliary quantity in
order to calculate the entanglement of formation \cite{bennet}.
Apart from the explicit formula (\ref{conc}), the concurrence of a mixed state
is defined as the minimum of the average concurrence 
$\langle c\rangle=\sum_i p_i c(\psi_i)$ over all 
decompositions $\rho=\sum_i p_i |\psi_i\rangle\langle\psi_i|$
of $\rho$ into pure states. After decomposing $\rho_s$ into product
states, also
the BSA, Eq. (\ref{lewsan}), defines a particular decomposition, and it
follows that
\begin{equation}
c(\rho)\leq (1-\lambda) c(\psi).
\label{concineq}
\end{equation}
This inequality implies
$c(\rho)+\lambda\leq 1$, which has already been conjectured in \cite{englert2}.
According to Theorem 2, equality in Eq.~(\ref{concineq}) holds if the BSA
of $\rho$ has full rank (in this case, $c(\psi)=1$). In other words:
the decomposition (\ref{lewsan}) is also optimal in the sense that it
minimizes the average concurrence. One might assume that this is true in
general, i.e. also in the second case, ${\rm rank}(\rho_s)=3$.
Indeed, there exist examples
where the inequality (\ref{concineq}) is saturated also in this case, e.g.
the generalized Werner states 
$\rho=x|\phi\rangle\langle\phi|+\frac{1-x}{4}{\mathbb{I}}$, with
$|\phi\rangle$ not maximally entangled. (The optimal decomposition of these
states is given in \cite{englert}.)
In general, however, we have found that the equality in (\ref{concineq})
does {\em not} always hold. Hence, the concurrence of $\rho$ and the
quantity $(1-\lambda) c(\psi)$
provide two inequivalent measures of entanglement. Indeed, it is 
straightforward to show that
$(1-\lambda) c(\psi)$ really {\em is} a good measure of entanglement, i.e.
it fulfills the following three conditions \cite{vedral}: it vanishes
if and only if $\rho$ is separable, is invariant under local unitary
operations and its expectation value is non-increasing under general local
operations (see \cite{vedral} for details).  

Before we present the proofs of Theorem 1 and 2, we want to demonstrate how 
to use the above results in order to construct the BSA for a given entangled
$\rho$ of rank $4$:
first, 
we calculate the smallest eigenvalue $\gamma$ and the corresponding
eigenvector $|\phi\rangle$ of the $4\times 4$ matrix $Y$, given by 
Eq. (\ref{y}).
(The eigenvalue $\gamma$ is not degenerate, see Lemma 7.) Then, we obtain
$|\psi\rangle$
from condition (i) of Theorem 1, $\lambda=1-2\sqrt{\gamma}$,
and $\rho_s$ from Eq. (\ref{lewsan}).
If $\rho_s$ is positive and separable, it is the BSA according to
Theorem 1. (It is not necessary to check $\rho_s^{T_B}|\phi\rangle=0$, since
this follows from the construction of $|\phi\rangle$, see Lemma 7 and
Lemmas 3-5.) 
If not, the BSA has rank 3, and from 
the second case of Theorem 1 we obtain the following set of
equations:
\begin{eqnarray}
[\rho|\tilde{\phi}\rangle\langle\tilde{\phi}|\rho]^{T_B}~|\phi\rangle & = &
\langle\tilde{\phi}|\rho|\tilde{\phi}\rangle\rho^{T_B}|\phi\rangle,
\label{eq1}\\
\nu|\tilde{\phi}\rangle\langle\tilde{\phi}|\rho|\tilde{\phi}\rangle\ +\ 
[|\phi\rangle\langle\phi|]^{T_B}\rho|\tilde{\phi}\rangle & = &
-\alpha\rho|\tilde{\phi}\rangle.
\label{eq2}
\end{eqnarray}
Here, we used $|\psi\rangle=\rho|\tilde{\phi}\rangle$ and
$(1-\lambda)^{-1}=\langle\tilde{\phi}|\rho|\tilde{\phi}\rangle$, see
Eqs. (\ref{lambda}) and (\ref{phitilde}) below.
These equations can be solved numerically for $|\tilde{\phi}\rangle$,
$|\phi\rangle$, $\alpha$, and $\nu$.
Possibly, there exist several solutions, but only one
with $\alpha,\nu\geq 0$ and which yields a positive and separable state
$\rho_s$ via Eq. (\ref{lewsan}). Thereby, we have found
the BSA of $\rho$ in a purely algebraic way, without employing any maximization
or optimization procedure.

Finally, we want to show how far this method can be used if
${\rm rank}(\rho)<4$, in particular if ${\rm rank}(\rho)=3$, since the case
${\rm rank}(\rho)=2$ has already been solved analytically\cite{englert}.
First, we note that any density matrix $\rho$ can be obtained
as a limit from the case of full rank.
Thereby, we obtain the following limiting case of Theorem 1
(the complete proof will be given below):  

\noindent {\bf Corollary.}
Let $\rho$ be an entangled state with ${\rm rank}(\rho)<4$.
$\rho=(1-\lambda)\left|\psi\right\rangle\left\langle\psi\right| + 
\lambda\rho_s$ is the optimal decomposition if and only if:

\begin{equation}
\exists_{|\phi\rangle} \rho_s^{T_B} |\phi\rangle=0,~c(\phi)>0,\ 
\exists_{|\tilde{\phi}\rangle} \rho_s |\tilde{\phi}\rangle=0,\ {\rm  and}\ 
\exists_{\alpha,\nu\geq 0} 
\left[\nu |\tilde{\phi}\rangle\langle\tilde{\phi}|+
\left[|\phi\rangle\langle\phi|\right]^{T_B}\right]\
|\psi\rangle=-\alpha |\psi\rangle.
\label{cor}
\end{equation}

Although Eq. (\ref{cor}) is nearly identical to case (ii) of Theorem 1, it
may also arise as a limit from case (i) (with $\nu=0$). Unlike in Theorem 1,
we must explicitly demand that $c(\phi)>0$ (in order to exclude solutions
which would not correspond to the optimal decomposition, see proof of the
corollary). 

For the solution of Eq. (\ref{cor}), the following observation
is helpful (cf. Lemma 1 of \cite{lewenstein}): if $|\tilde{\phi}\rangle$ is
not in the kernel of $\rho$, then
$|\psi\rangle=\rho|\tilde{\phi}\rangle$ and $\lambda$ is given by
Eq. (\ref{lambda}) (well defined, since $|\psi\rangle$ is in the range
of $\rho$). Consequently, Eq. (\ref{cor}) reduces to
Eqs. (\ref{eq1},\ref{eq2}), as in the case of full rank (with the
additional constraint $c(\phi)>0$). Hence, the recipe for obtaining the BSA
of $\rho$ with ${\rm rank}(\rho)=3$
is as follows: first, try to find a solution of Eqs. (\ref{eq1},\ref{eq2})
with $\alpha,\nu\geq 0$ and $c(\phi)>0$ which yieds a positive and separable
$\rho_s$. If such a solution does not exist, we know that
$|\tilde{\phi}\rangle$ must be in the kernel of $\rho$, which uniquely
determines $|\tilde{\phi}\rangle$ (since we assumed ${\rm rank}(\rho)=3$). 
Then, Eq. (\ref{cor}) can be solved numerically for
$|\psi\rangle$, $|\phi\rangle$, $\alpha$, $\nu$ and $\lambda$
(after inserting $|\tilde{\phi}\rangle$ and replacing $\rho_s^{T_B}$ by
$\rho^{T_B}-(1-\lambda)[|\psi\rangle\langle
\psi|]^{T_B}$). Furthermore, the following fact may be useful:
if the kernel of $\rho$ contains a product vector $|e,f\rangle$, then 
the solution of Eq. (\ref{cor}) fulfills
$|\tilde{\phi}\rangle\perp|e,f\rangle$ and $|\phi\rangle\perp|e,f^*\rangle$
(see proof of the corollary).

\subsection{Proof of Theorem 1}

Let $\rho_s$, as given by Eq. (\ref{lewsan}), be the BSA of $\rho$. The
maximality condition for $\lambda$ and the uniqueness of the BSA
\cite{lewenstein} imply \cite{englert}:
\begin{itemize}
\item[(a)]the state $\rho_s+\epsilon|\psi\rangle\langle\psi|$ is
non-separable
for $\epsilon>0$, and
\item[(b)]the state $\rho-(1-\lambda)|\psi'\rangle\langle\psi'|$
is either non-separable or non-positive for each $\psi'\neq\psi$.
\end{itemize}
In order to compact the notation we shall use in the following the notation
$\mu=1-\lambda$.

According to the Peres-Horodecki criterion of separability
\cite{peres,horodecki3}, condition (a) implies:
\begin{equation}
\forall_{\epsilon>0}
\exists_{\left|\phi_{\epsilon}\right\rangle}
\left\langle\phi_{\epsilon}\right|\left.\rho_s^{T_B}+
\epsilon\left[\left|\psi\right\rangle\left\langle\psi\right|\right]^{T_B}
\right.\left|\phi_{\epsilon}\right\rangle<0.
\label{phieps}
\end{equation}
On the other hand, since $\rho_s$ is separable, the same criterion
establishes
the positivity of $\rho_s^{T_B}$. Thus, from (\ref{phieps}) and the
continuity
argument, there is such $\phi$ that
\begin{equation}
\rho_s^{T_B}\left|\phi\right\rangle=0.
\label{ev2}
\end{equation}
Since we assumed ${\rm rank}(\rho)=4$ and the rank of a projection is one,
the
rank of $\rho_s$ must be at least three. Then ${\rm rank}(\rho_s^{T_B})=3$,
as a consequence of Lemma 1 (c.f.\ Appendix A).

Now we exploit condition (b). Let's consider
\begin{equation}
\lambda\rho_s^{\prime}:=\rho-\mu\left|\psi^{\prime}
\right\rangle\left\langle\psi^{\prime}\right|,
\end{equation}
with
\begin{equation}
\left|\psi^{\prime}\right\rangle=\frac{\left|\psi\right\rangle+
\epsilon \left|\Delta\psi\right\rangle}{\sqrt{1+\epsilon^2}},
\end{equation}
where $\langle\Delta\psi|\Delta\psi\rangle=1$ and
 $\left\langle\psi\right.\left|\Delta\psi\right\rangle=0$.
(Obviously, any pure state can be written in this form.)

To the two lowest orders in $\epsilon$ we have:
\begin{equation}
\lambda\rho_s^{\prime}=\lambda\rho_s- \mu \left(\epsilon \left|\psi
\right\rangle\left\langle\Delta\psi\right|+\epsilon \left|\Delta\psi
\right\rangle\left\langle\psi\right|+\epsilon^2
|\Delta\psi\rangle\langle\Delta\psi|-\epsilon^2
|\psi\rangle\langle\psi|\right).
\label{rhosp}
\end{equation}

In the following, we consider separately two cases of different ranks of
$\rho$.
\begin{itemize}
\item[(i)]${\rm rank}(\rho_s)=4$

Then, for $\epsilon$ small enough,  $\rho_s^{\prime}$  is positive definite
for
each $|\Delta\psi\rangle$. According to the optimality condition (b) above,
$\rho_s^{\prime}$ must be non-separable, i.e.\ there exists such
$\left|\phi^{\prime}\right\rangle$ that
$\left\langle\phi^{\prime}\right|\left.\rho_s^{\prime T_B}\right.\left|
\phi^{\prime}\right\rangle< 0$.

Since $\rho_s^{T_B}$ has rank $3$,
\begin{equation}
\left|\phi^{\prime}\right\rangle=\left|\phi\right\rangle
+\left|\Delta\phi\right\rangle,
\end{equation}
with $|\Delta\phi\rangle\to 0$ if $\epsilon\to 0$.
Now from (\ref{ev2}) we
obtain, to the first order in $\epsilon$:
\begin{equation}
\left\langle\Delta\phi\right|\lambda\rho_s^{T_B}\left|\Delta\phi\right\rangle
-\mu\epsilon\langle\phi|~\left[\left|\psi
\right\rangle\left\langle\Delta\psi\right|+\left|\Delta\psi
\right\rangle\left\langle\psi\right|\right]^{T_B}~|\phi\rangle\leq
0.
\label{neg1}
\end{equation}
But $\rho_s$ is, by assumption, separable; consequently $\rho_s^{T_B}$ is
positive definite
\begin{equation}
\left\langle\Delta\phi\right|\lambda\rho_s^{T_B}\left|\Delta\phi\right\rangle
\geq 0,
\end{equation}
and (\ref{neg1}) implies
\begin{equation}
\langle\phi|~\left[\left|\psi
\right\rangle\left\langle\Delta\psi\right|+\left|\Delta\psi
\right\rangle\left\langle\psi\right|\right]^{T_B}~|\phi\rangle\geq 0,
\end{equation}
which can be equivalently written as
\begin{equation}
\text{Tr}\left\{\left[\left|\psi
\right\rangle\left\langle\Delta\psi\right|+\left|\Delta\psi
\right\rangle\left\langle\psi\right|\right]^{T_B}
\left|\phi\right\rangle\left\langle\phi\right|\right\}\geq 0.
\label{tr1}
\end{equation}
For arbitrary operators $A$ and $B$ we have
$\text{Tr}A^{T_B}B=\text{Tr}AB^{T_B}$, thus from (\ref{tr1}) we obtain
\begin{equation}
\text{Tr}\left\{\left[\left|\psi
\right\rangle\left\langle\Delta\psi\right|+\left|\Delta\psi
\right\rangle\left\langle\psi\right|\right]
[\left|\phi\right\rangle\left\langle\phi\right|]^{T_B}\right\}\geq 0.
\label{tr2}
\end{equation}
This, however, is
equivalent to
\begin{equation}
\langle\Delta\psi|~[\left|\phi\right\rangle\left\langle\phi\right|
]^{T_B}~|\psi\rangle+
\langle\psi|~[\left|\phi\right\rangle\left\langle\phi\right|]^{T_B}~\left|
\Delta\psi\right\rangle\geq 0
\label{perp1}
\end{equation}
for all $|\Delta\psi\rangle\perp|\psi\rangle$. Since (\ref{perp1}) is linear in
$\left|\Delta\psi\right\rangle$, changing $\left|\Delta\psi\right\rangle$
into
$-\left|\Delta\psi\right\rangle$ reverses the inequality, hence in fact it
must be that
\begin{equation}
\langle\Delta\psi|~[\left|\phi\right\rangle\left\langle\phi\right|
]^
{T_B}~|\psi\rangle=0.
\label{cond1}
\end{equation}
The above equality must be fulfilled by all
$|\Delta\psi\rangle\perp|\psi\rangle$. 
This is
possible only if $\left|\psi\right\rangle$ is an eigenvector of
$A=[\left|\phi\right\rangle\left\langle\phi\right|]^{T_B}$. 
(Note that, although $T_B$ and $|\phi\rangle$ depend on the local basis
of ${\mathcal H}_B$, the operator $A$ is basis-independent, i.e. transforms
in the usual way, Eq. (\ref{transf2}), under local unitary transformations.) 
To
arrive at the first case (i) of Theorem 1, it remains to be shown that
the sign of the corresponding eigenvalue $\alpha$ is negative. This, however,
follows
from the limit $\epsilon\to 0$ of Eq. (\ref{phieps})
\begin{equation}
\langle\phi|~[|\psi\rangle\langle\psi|]^{T_B}~|\phi\rangle\leq 0,
\end{equation}
after using again the identity ${\rm Tr}A^{T_B}B={\rm Tr}AB^{T_B}$.
Furthermore, $\alpha$ cannot be zero - otherwise (according to Lemma 2)
$|\phi\rangle$
would be a separable, i.e.\ a product state: $\phi=|e,f\rangle$,
and since $\rho_s^{T_B}|e,f\rangle=0$ (c.f. (\ref{ev2})), we have
$\rho_s|e,f^*\rangle=0$, which contradicts the assumption
${\rm rank}(\rho_s)=4$. From Lemma 2 (c.f.\ Appendix) we infer that
$|\psi\rangle$ is maximally
entangled. This provides an alternative proof of the fact proved in
\cite{karnas} that if $\rho$ and $\rho_s$ are of
maximal rank then $|\psi\rangle$ in (\ref{lewsan}) is maximally entangled.

\item[(ii)]Second case: ${\rm rank}(\rho_s)<4$

We assumed that $\rho$ has rank 4, so ${\rm rank}(\rho_s)=3$.
From Lemma 1 in \cite{lewenstein}, we know that
\begin{equation}
1-\lambda=\frac{1}{\langle\psi|\rho^{-1}|\psi\rangle}.
\label{lambda}
\end{equation}
Furthermore, it is easy to check that
\begin{equation}
|\tilde{\phi}\rangle=\rho^{-1}|\psi\rangle
\label{phitilde}
\end{equation}
fulfills $\rho_s|\tilde{\phi}\rangle=0$.
Since ${\rm rank}(\rho_s)=3$,
$\rho_s'$, Eq.(\ref{rhosp}), is
positive definite if
\begin{equation}
\langle\tilde{\phi}|\ (\epsilon|\psi\rangle\langle\Delta\psi| +
\epsilon|\Delta\psi\rangle\langle\psi|
+\epsilon^2
|\Delta\psi\rangle\langle\Delta\psi|-\epsilon^2
|\psi\rangle\langle\psi|
)\ |\tilde{\phi}\rangle<0.
\label{pos}
\end{equation}
Obviously, this condition is fulfilled if
$|\Delta\psi\rangle\perp|\tilde{\phi}\rangle$.
[$\langle\psi|\tilde{\phi}\rangle\neq 0$ follows
from Eq. (\ref{lambda}).] Hence
(as in case 1), all such $|\Delta\psi\rangle$ must fulfill Eq. (\ref{cond1}).

This is equivalent to
\begin{equation}
(1-|\psi\rangle\langle\psi|- |\tilde{\psi}\rangle\langle
\tilde{\psi}|)\,[|\phi\rangle\langle\phi|]^{T_B}\ |\psi\rangle=0,
\label{cond2}
\end{equation}
where $|\tilde{\psi}\rangle$ is defined such that
$|\tilde{\psi}\rangle\perp|\psi\rangle$ and $|\tilde{\psi}\rangle$ 
and $|\psi\rangle$ span the
same two-dimensional subspace as $|\tilde{\phi}\rangle$ and $|\psi\rangle$. (We
assume that $|\psi\rangle\neq|\tilde{\phi}\rangle$; otherwise, $\rho_s'$ is
positive for all $|\Delta\psi\rangle$, and we get the same result as in the
first case, which below will turn out to be a special case of the result in the
second case.)

We still have to check the case $|\Delta\psi\rangle=|\tilde{\psi}\rangle$.
Then, it is always possible to multiply $|\Delta\psi\rangle$ by a phase
factor such that $\rho_s'$ is positive, see Eq. (\ref{pos})
in first order of $\epsilon$. This leads
us (as in case 1) to Eq. (\ref{perp1}). It follows that
\begin{equation}
-\nu\
\langle\tilde{\psi}|\tilde{\phi}\rangle\langle\tilde{\phi}|\psi\rangle\
=\ \langle\tilde{\psi}|~[|\phi\rangle\langle\phi|]^{T_B}~|\psi\rangle
\label{perp2}
\end{equation}
with a nonnegative real parameter $\nu$. Otherwise, $|\Delta\psi\rangle$ could
be multiplied by a phase factor such that Eq. (\ref{pos}) is fulfilled and Eq.
(\ref{perp1}) not. (Note that $\langle\tilde{\psi}|\tilde{\phi}
\rangle\langle\tilde{\phi}|\psi\rangle\neq 0$, since
$\langle\psi|\tilde{\phi}\rangle\neq 0$ follows from Eq. (\ref{lambda}), and
$\langle\tilde{\psi}|\tilde{\phi}\rangle\neq 0$ from the construction of
$\tilde{\psi}$.)

The two conditions Eqs. (\ref{perp2},\ref{cond2}) are equivalent to the
following condition: $|\psi\rangle$ is an eigenvector of the operator
\begin{equation}
A=\nu |\tilde{\phi}\rangle\langle\tilde{\phi}|
+[|\phi\rangle\langle\phi|]^{T_B}.
\end{equation}
To complete the first part of the proof
of Theorem 1, we will show now that the corresponding eigenvalue $\alpha$
cannot be
positive.

As a consequence of Lemma 2, $A$ has at least three nonnegative eigenvalues.
However, there is also at least one nonpositive eigenvalue.
This follows from the existence of a product
vector $|e,f\rangle\in R(\rho_s)$ such that $|e,f^*\rangle\in R(\rho_s^{T_B})$,
as shown in \cite{karnas}, which implies $\langle e,f|A|e,f\rangle=0$.
Furthermore, $A$ cannot have more than one zero eigenvalue:
otherwise, $|\phi\rangle$ would have to be a product vector (see Lemma 2),
and $|\tilde\phi\rangle$
would be the corresponding partially transposed product vector. Hence,
$|\tilde{\phi}\rangle\langle\tilde{\phi}|$ and
$[|\phi\rangle\langle\phi|]^{T_B}$ would be identical and proportional
to $A$, and
$|\psi\rangle$, as an {\em entangled} eigenvector of $A$, would have to be
perpendicular to $|\tilde{\phi}\rangle$, i.e. ${\rm rank}(\rho)=3$,
which contradicts the assumption ${\rm rank}(\rho)=4$.

The above considerations about the spectrum of $A$ are useful for the
following reason: let us assume that there exists an entangled state $\rho'$
with $\alpha'<0$ which has the property that $\rho(x)=x\rho+(1-x)\rho'$ is
entangled for $x\in[0,1]$. ($\rho'$ may be a state with BSA of rank 4,
for which we have already shown above that $\alpha'<0$.)
Now, the optimal
decomposition (\ref{lewsan}) - in particular the eigenvalue $\alpha(x)$ -
changes smoothly when varying $x$ from $0$ to $1$
(this follows from the uniqueness of the optimal decomposition).
Since, as shown above, $A$ (having one nonpositive and three nonnegative
eigenvalues) cannot have two zero eigenvalues, a crossing of
eigenvalues at zero is not possible, and $\alpha=\alpha(1)\leq 0$ follows
from $\alpha'=\alpha(0)<0$. 

It remains to be shown that a state $\rho'$ with the above properties
exists. For this purpose, we consider the Werner states
$\rho'=y|\psi'\rangle\langle\psi'|+\frac{1-y}{4}\mathbb{I}$, with
maximally entangled $|\psi'\rangle$. For these states, it has been shown
in \cite{englert} that the pure state in the optimal decomposition equals
$|\psi'\rangle$ and $\lambda'=3(1-y)/2$. It follows that 
${\rm rank}(\rho_s')=4$, and $\alpha'<0$, as shown above (first case).
Now, we choose $|\psi'\rangle$ as the eigenvector of
$[|\chi\rangle\langle\chi|]^{T_B}$ with negative eigenvalue (such an 
eigenvalue exists according to Lemma 2), where $|\chi\rangle$ is an entangled
pure state with $\langle\chi|\rho^{T_B}|\chi\rangle<0$ (exists, since
$\rho$ is entangled). Using $\langle\psi'|~[|\chi\rangle\langle\chi|]^{T_B}~
|\psi'\rangle=\langle\chi|~[|\psi'\rangle\langle\psi'|]^{T_B}~
|\chi\rangle$, it follows that 
$\langle\chi|(\rho')^{T_B}|\chi\rangle<0$ for large enough $y$, hence 
also $\langle\chi|\rho(x)^{T_B}|\chi\rangle<0$ for $x\in [0,1]$, i.e.
$\rho(x)$ is entangled.
\end{itemize}

Finally, we will prove the reverse direction of Theorem~1, i.e. that both cases
(i) and (ii) are also {\em sufficient} for the optimality of the decomposition
(\ref{lewsan}). For this purpose, let us assume that there exists another
decomposition with larger $\lambda$. Then, because of the convexity of the set
of separable states, such a decomposition with larger $\lambda$ also exists in
the infinitesimal neighborhood of $\{\lambda,|\psi\rangle\}$. Hence, for each
(infinitesimal small) $\epsilon>0$, there exists $\lambda'=\lambda+
\Delta\lambda$ (with $\Delta\lambda>0$ and $\Delta\lambda\to 0$ if
$\epsilon\to 0$) and 
$|\Delta\psi\rangle\perp|\psi\rangle$
such that
\begin{equation}
\lambda' \rho_s^{\prime}=\lambda\rho_s+\Delta\lambda|\psi\rangle\langle\psi|-
(1-\lambda') \left(\epsilon \left|\psi
\right\rangle\left\langle\Delta\psi\right|+\epsilon \left|\Delta\psi
\right\rangle\left\langle\psi\right|+\epsilon^2
|\Delta\psi\rangle\langle\Delta\psi|-\epsilon^2
|\psi\rangle\langle\psi|\right)
\label{rhosp2}
\end{equation}
is separable. Now, let us assume that there exists $|\phi\rangle$ with
$\rho_s^{T_B}|\phi\rangle=0$ and either condition (i) or (ii) from
Theorem (i) is fulfilled.
In the following, we will show that both (i) or (ii) lead to a
contradiction, since either $\langle\phi|\rho_s^{\prime T_B}|\phi\rangle<0$ or
$\langle\tilde{\phi}|\rho_s^{\prime}|\tilde{\phi}\rangle<0$.

\begin{itemize}
\item[(i)] implies $\langle\Delta\psi|~[|\phi\rangle\langle\phi|]^{T_B}~
|\psi\rangle=0$, $\langle\psi|~[|\phi\rangle\langle\phi|]^{T_B}~
|\psi\rangle<0$, and
 $\langle\Delta\psi|~[|\phi\rangle\langle\phi|]^{T_B}~
|\Delta\psi\rangle>0$. (The third inequality follows from the spectrum of
$[|\phi\rangle\langle\phi|]^{T_B}$, see Lemma 2.)
Inserting into Eq. (\ref{rhosp2}) immediately
yields $\langle\phi|\rho_s^{\prime T_B}|\phi\rangle<0$.

\item[(ii)] implies $\langle\psi|~[|\phi\rangle\langle\phi|]^{T_B}~
|\psi\rangle=\alpha-\nu \langle\psi|\tilde{\phi}\rangle\langle\tilde{\phi}
|\psi\rangle$ and $\langle\Delta\psi|~[|\phi\rangle\langle\phi|]^{T_B}~
|\psi\rangle=-\nu \langle\Delta\psi|\tilde{\phi}\rangle\langle\tilde{\phi}
|\psi\rangle$. Inserting into Eq. (\ref{rhosp2}) yields:
\begin{equation}
\langle\phi|\rho_s^{\prime T_B}|\phi\rangle = \Delta\lambda\
\alpha+(1-\lambda')
\epsilon^2 (\alpha-\beta)-\nu \langle\tilde{\phi}|\rho_s^\prime
|\tilde{\phi}\rangle,
\end{equation}
where $\beta=\langle\Delta\psi|A|\Delta\psi\rangle$. Since $\alpha\leq 0$ and
$\alpha<\beta$ (remember that $A=\nu |\tilde{\phi}\rangle\langle\tilde{\phi}|
+[|\phi\rangle\langle\phi|]^{T_B}$ has three nonnegative eigenvalues, i.e.
$\alpha$ is the smallest eigenvalue of $A$), it follows that $\rho_s^{\prime}$
is either non-positive or non-separable.
\end{itemize}
$\Box$

\subsection{Proof of Theorem 2}

Let us assume that Eq. (\ref{lewsan}) is the optimal decomposition of
$\rho$, with $\rho_s$ of rank 4. According to Theorem 1 (and Lemma 2), we
know that $|\psi\rangle$ is maximally entangled, i.e.
$c(\psi)=1$. Hence, we can use Lemma~3 of Appendix A to write
\[
\lambda\rho_s^{T_B}=\rho^{T_B}-\mu[|\psi\rangle\langle\psi|]^{T_B}
=\rho^{T_B}-\mu\left(\frac{1}{2}\mathbb{I}-
|\widetilde{\psi}\rangle\langle\widetilde{\psi}|\right),
\]
where $|\tilde{\psi}\rangle$ is defined by
\begin{equation}
[|\psi\rangle\langle\psi|]^{T_B}|\widetilde{\psi}\rangle
=-\frac{1}{2}|\widetilde{\psi}\rangle.
\label{psitilde}
\end{equation}
Consequently, for an arbitrary $|\phi^{\prime}\rangle$
\begin{equation}
0\leq\lambda\langle\phi^{\prime}|\rho_s^{T_B}|\phi^{\prime}\rangle
=\langle\phi^{\prime}|\rho^{T_B}|\phi^{\prime}\rangle
+\mu|\langle\phi^{\prime}|\widetilde{\psi}\rangle|^2
-\frac{\mu}{2}.
\label{cphip}
\end{equation}
For $|\phi^{\prime}\rangle=|\phi\rangle$, the above equation, due to
(\ref{ev2}), reads
\begin{equation}
0=\langle\phi|\rho^{T_B}|\phi\rangle
+\mu|\langle\phi|\widetilde{\psi}\rangle|^2
-\frac{\mu}{2}.
\label{cphi1}
\end{equation}
Observe now that because of (i) (Theorem 1) and (\ref{psitilde}), we can
apply Lemma
5, concluding that $|\phi\rangle$ and
$|\widetilde{\psi}\rangle$ have a common Schmidt basis, hence,
according to Lemma 4 we can
rewrite (\ref{cphi1}) as
\begin{equation}
0=\langle\phi|\rho^{T_B}|\phi\rangle+\frac{\mu}{2}c(\phi).
\label{cphi2}
\end{equation}
Using the results of the same Lemma we can estimate the last two terms on the
right-hand side of (\ref{cphip}) by $\frac{\mu}{2} c(\phi^{\prime})$:
\begin{equation}
0\leq\langle\phi^{\prime}|\rho^{T_B}|\phi^{\prime}\rangle+
\frac{\mu}{2}c(\phi^{\prime}).
\label{cphip1}
\end{equation}

In order to simplify equations we are going to consider below, let us make the
following observation. Both equation (\ref{cphi2}) and inequality 
(\ref{cphip1})
are bilinear in $|\phi\rangle$ if only we calculate the concurrence according
to (\ref{purec}) regardless of the normalization of $|\psi\rangle$. Obviously
such a quantity is not limited from above, but this will not play any role in
the following. The final formula will involve only normalized vectors.

Substituting $|\phi^{\prime}\rangle=|\phi\rangle+\varepsilon|\Delta\phi\rangle$
(with arbitrary $\epsilon$ and $|\Delta\phi\rangle$)
to (\ref{cphip1}) and using (\ref{cphi2}), we obtain in the lowest order in
$\varepsilon$
\begin{eqnarray}
0\leq\varepsilon\left(\langle\Delta\phi|\rho^{T_B}|\phi\rangle
+\langle\phi|\rho^{T_B}|\Delta\phi\rangle+\frac{\mu}{2}
\frac{dc(\phi+\varepsilon\Delta\phi)}{d\varepsilon}|_{\varepsilon=0}\right).
\label{bound1}
\end{eqnarray}
From the definition of concurrence (\ref{purec}) we obtain
\[
c(\phi+\varepsilon\Delta\phi)=c(\phi)
+\varepsilon\left(\langle\Delta\phi|\Sigma|\phi^*\rangle
+\langle\phi^*|\Sigma|\Delta\phi\rangle\right)+o(\varepsilon^2),
\]
after adjusting the phase of $|\phi\rangle$ to make
$\langle\phi|\Sigma|\phi^*\rangle$ real and positive and using
$\langle\Delta\phi|\Sigma|\phi^*\rangle=
\langle\phi|\Sigma|\Delta\phi^*\rangle=
\langle\Delta\phi^*|\Sigma|\phi\rangle^*=
\langle\phi^*|\Sigma|\Delta\phi\rangle^*$ which is a consequence of
$\Sigma=\Sigma^\dagger=\Sigma^*$. Thus, we can rewrite (\ref{bound1})
as
\begin{equation}
\langle\Delta\phi|\rho^{T_B}|\phi\rangle
+\langle\phi|\rho^{T_B}|\Delta\phi\rangle
+\frac{\mu}{2}\left(\langle\Delta\phi|\Sigma|\phi^*\rangle
+\langle\phi^*|\Sigma|\Delta\phi\rangle\right)\geq 0,
\label{bound2}
\end{equation}
valid for an arbitrary $|\Delta\phi\rangle$. Again, considering
(\ref{bound2})
for $|\Delta\phi\rangle$ and $-|\Delta\phi\rangle$, we conclude that in fact
(\ref{bound2})
is an equality
\[
\langle\Delta\phi|\Psi\rangle+\langle\Psi|\Delta\phi\rangle=0,
\]
where
\[
|\Psi\rangle=\rho^{T_B}|\phi\rangle+\frac{\mu}{2}\Sigma|\phi^*\rangle.
\]
Since $|\Delta\phi\rangle$ is arbitrary, we have $|\Psi\rangle=0$ and,
consequently
\begin{equation}
\rho^{T_B}|\phi\rangle=-\frac{\mu}{2}\Sigma|\phi^*\rangle.
\label{main1}
\end{equation}
Short manipulations using $\Sigma^2=1$ allow for rewriting (\ref{main1}) as an
eigenvalue equation
\begin{equation}
\Sigma(\rho^{T_B})^*\Sigma\rho^{T_B}|\phi\rangle=\frac{\mu^2}{4}|\phi\rangle
.
\label{main}
\end{equation}
In Appendix B (Lemma 6),
we show that the smallest eigenvalue $\gamma$ of
$Y=\Sigma(\rho^{T_B})^*\Sigma\rho^{T_B}$ is
given by $\gamma=c^2(\rho)/4$, where $c(\rho)$ is the concurrence
of $\rho$. Furthermore, it follows from Lemma 7 that
$\mu^2/4$ is the smallest eigenvalue of $Y$, since
$\langle\phi|\rho^{T_B}|\phi\rangle<0$ according to Eq. (\ref{main1}). $\Box$

\subsection{Proof of the Corollary}

Let $\rho$ be an entangled state with ${\rm rank}(\rho)<4$,
and $\rho=(1-\lambda)|\psi\rangle\langle\psi|+\lambda\rho_s$ its
optimal decomposition.
Furthermore, we define $\rho_\epsilon:=(1-\epsilon)\rho+\epsilon
{\mathbb{I}}/4$ (where $\mathbb{I}$ is the $4\times 4$ identity operator).
Obviously, ${\rm rank}(\rho_\epsilon)=4$ and
$\rho=\lim_{\epsilon\to 0}\rho_\epsilon$. Since the optimal decomposition
of $\rho_\epsilon$ varies continously with $\epsilon$, it follows from
Theorem 1 in the limit $\epsilon\to 0$ that
\begin{equation}
\exists_{|\phi\rangle} \rho_s^{T_B} |\phi\rangle=0,\ 
\exists_{|\tilde{\phi}\rangle} \rho_s |\tilde{\phi}\rangle=0,\ {\rm  and}\ 
\exists_{\alpha,\nu\geq 0} 
\left[\nu |\tilde{\phi}\rangle\langle\tilde{\phi}|+
\left[|\phi\rangle\langle\phi|\right]^{T_B}\right]\
|\psi\rangle=-\alpha |\psi\rangle.
\label{cor2}
\end{equation}
This equation includes both cases (i) and (ii) of Theorem 1. 
(In the former case, $\nu=0$ and $|\tilde{\phi}\rangle$ is an element of
the kernel of $\rho$.) On the other hand, if we assume that
$|\phi\rangle$ is not a product vector (i.e. $c(\phi)>0$), then  
Eq. (\ref{cor2}) is 
sufficient for the optimality of the decomposition. This can be proven in the
same way as above in the proof of the
reverse direction of Theorem 1. Indeed, no
assumption about the rank of $\rho$ is needed there, except for showing that
$|\phi\rangle$ is not a product vector (which ensures that
$-\alpha$ is strictly the smallest eigenvalue of  
$A=\nu |\tilde{\phi}\rangle\langle\tilde{\phi}|
+[|\phi\rangle\langle\phi|]^{T_B}$).
It remains to be shown that $c(\phi)>0$ is necessary for the optimality of the
decomposition.
For this purpose, let us assume that $|\phi\rangle=|e,f\rangle$ is a
product vector. From the definition of
partial transposition, we know that $\rho_s|e,f^*\rangle=0$, and from
Eq. (\ref{cor2}) with $\alpha\geq 0$ that
$|\psi\rangle\perp|e,f^*\rangle$. This implies $\rho |e,f^*\rangle=0$.
Then, it is easy to show that
the BSA of $\rho_\delta:=(1-\delta)\rho+\delta |e,f^*\rangle\langle e,f^*|$
is given by ${\rho_\delta}_s=(1-\delta)\rho_s+\delta
|e,f^*\rangle\langle e,f^*|$.
Hence, any vector $|\phi_\delta\rangle$ with ${\rho_\delta}_s^{T_B}
|\phi_\delta\rangle =0$ fulfills $|\phi_\delta\rangle\perp|e,f\rangle$, which,
in the limit $\delta\to 0$, contradicts
the assumption $|\phi\rangle=|e,f\rangle$. $\Box$

\acknowledgments
We are grateful for discussions with Karol \.Zyczkowski, Berthold-Georg
Englert, and Pawe{\l}~Horodecki. 
M.K. was supported by Polish Komitet Bada\'n Naukowych through
research Grant No 2 P03B 072 19, and T.W. by
DAAD through a HSP III Kurzstipendium f\"ur Doktoranden. 

\appendix
\section{}

In the Appendix we formulate and prove several Lemmas used in the main part
of the paper.

\noindent{\bf Lemma 1.} Let $\rho_s$ be a two qubit density matrix.
If ${\rm rank}(\rho_s^{T_B})\leq 2$, then ${\rm rank}(\rho_s)=
{\rm rank}(\rho_s^{T_B})$.

\noindent{\bf Proof:} Since every two-dimensional subspace contains a
product vector
\cite{lewenstein}, also the kernel of $\rho_s^{T_B}$ must do so, i.e.
$\rho_s^{T_B}|e,f\rangle=0$. It follows that $\rho_s|e,f^*\rangle=0$.
Indeed, from (\ref{efref}) we have
$\langle e,f^*|\rho|e,f^*\rangle=\langle e,f|\rho^{T_B}|e,f\rangle=0$, and
since
$\rho_s$ as a density matrix is positive definite, $\rho|e,f^*\rangle=0$. By
local unitary transformations in both subspaces we can choose
$|e,f^*\rangle=|0,0\rangle=|e,f\rangle$. Equations $\rho|0,0\rangle=0$ and
$\rho^{T_B}|0,0\rangle=0$ together with hermiticity of both matrices leave
only
six nonvanishing elements in each of them, and by inspection one checks that
their characteristic polynomials (hence also the spectra) are identical.

\noindent{\bf Lemma 2.} For an arbitrary
$|\phi\rangle$ the matrix $[|\phi\rangle\langle\phi|]^{T_B}$ has eigenvalues
\[
-\frac{c}{2},\frac{c}{2},\frac{1-\sqrt{1-c^2}}{2},\frac{1+\sqrt{1-c^2}}{2},
\]
where $c=c(\phi)$ is the concurrence of $|\phi\rangle$.
If $c>0$, the eigenvector belonging to the negative eigenvalue is maximally
entangled.

\noindent{\bf Proof:} The first part of the Lemma is proven by an explicit
calculation. In order to prove the second statement,
let $L=U\otimes V$ be a local transformation, and
$|\phi^{\prime}\rangle=L|\phi\rangle$. Then
\[
[|\phi^{\prime}\rangle\langle\phi^{\prime}|]^{T_B}
=L^{\prime}[|\phi\rangle\langle\phi|]^{T_B}L^{\prime\dagger},
\]
where $L^{\prime}=U\otimes V^*$. Observe that $L^{\prime}$ is a
local transformation, hence it does not influence the concurrence of
vectors.
Now,
\begin{equation}
[|\phi\rangle\langle\phi|]^{T_B}|\psi\rangle=-\frac{c(\phi)}{2}|\psi\rangle
\Leftrightarrow
[|\phi^{\prime}\rangle\langle\phi^{\prime}|]^{T_B}|\psi^{\prime}
\rangle=
-\frac{c(\phi)}{2}|\psi^{\prime}\rangle,
\label{l5aa}
\end{equation}
where $|\psi^{\prime}\rangle=L^{\prime}|\psi\rangle$. Let us now
choose $L$
such that it brings $|\phi\rangle$ to its Schmidt basis:
\[
|\phi^{\prime}\rangle=L|\phi\rangle=\left[
\begin{array}{c}
\lambda_1 \\
0 \\
0 \\
\lambda_2
\end{array}
\right].
\]
It is now straightforward to show that $|\psi^{\prime}\rangle$ in
(\ref{l5aa}) has the form
\begin{equation}
|\psi^{\prime}\rangle=\frac{1}{\sqrt{2}}e^{i\delta}
\left[
\begin{array}{c}
0 \\
1 \\
-1\\
0
\end{array}
\right],
\end{equation}
hence $|\psi^{\prime}\rangle$ is maximally entangled and the same is true
about
$|\psi\rangle$ which is obtained from $|\psi^{\prime}\rangle$ by a local
transformation $L^{\prime}$.

(Similar versions of Lemma 1 and Lemma 2 can also be found in \cite{sanpera}.)

\noindent{\bf Lemma 3.} If $|\psi\rangle$ is maximally entangled then
\begin{equation}
[|\psi\rangle\langle\psi|]^{T_B}=\frac{1}{2}\mathbb{I}-
|\widetilde{\psi}\rangle\langle\widetilde{\psi}|,
\label{l3a}
\end{equation}
where $\mathbb{I}$ is the $4\times 4$ identity operator and
$|\widetilde{\psi}\rangle$ is the eigenvector of
$[|\psi\rangle\langle\psi|]^{T_B}$ with the negative eigenvalue i.e.\
\begin{equation}
[|\psi\rangle\langle\psi|]^{T_B}|\widetilde{\psi}\rangle=-\frac{1}{2}
|\widetilde{\psi}\rangle.
\label{l3b}
\end{equation}
According to Lemma 2, $|\widetilde{\psi}\rangle$ is maximally entangled.

\noindent{\bf Proof:}
Since $[|\psi\rangle\langle\psi|]^{T_B}$ is Hermitian, it has, in addition to
$|\psi_4\rangle:=|\widetilde{\psi}\rangle$ three other orthogonal
eigenvectors
$|\psi_i\rangle$, $i=1,2,3$ fulfilling, according to Lemma 2
\begin{equation}
[|\psi\rangle\langle\psi|]^{T_B}|\psi_i\rangle=\frac{1}{2}|\psi_i\rangle,
\quad
i=1,2,3.
\label{l3c}
\end{equation}
Using (\ref{l3c}) and (\ref{l3b}) together with the orthonormality of the
eigenvectors, $\langle\psi_i|\psi_j\rangle=\delta_{ij}$, $i=1,2,3,4$, one
sees
that the actions of both sides of (\ref{l3a}) give the same results on the
complete orthonormal set $|\psi_i\rangle$, $i=1,2,3,4$, which establishes
the
(\ref{l3a}) as a matrix equation.

\noindent{\bf Lemma 4.} For arbitrary $|\phi\rangle$,
\begin{equation}
\max_{m.e.}|\langle\phi|\psi\rangle|^2=\frac{1}{2}+\frac{1}{2}c(\phi),
\label{l4}
\end{equation}
where the maximum is taken over all maximally entangled $|\psi\rangle$. The
maximum is attained if $|\psi\rangle$ and $|\phi\rangle$ have a common
Schmidt basis.

\noindent{\bf Proof:} By a local unitary transformation (which does not
change
neither $|\langle\phi|\psi\rangle|^2$ nor the entanglements of
$|\phi\rangle$
and $|\psi\rangle$)  we can bring $|\phi\rangle$ to its Schmidt basis:
\[
|\phi\rangle=
\left[
\begin{array}{c}
\lambda_1 \\
0 \\
0 \\
\lambda_2
\end{array}
\right], \quad \lambda_i\geq 0, \quad \lambda_1^2+\lambda_2^2=1.
\]
Using the general form (\ref{megen}) of a maximally entangled state, we
conclude
that in the new basis
\[
|\langle\phi|\psi\rangle|^2=
|a_1\lambda_1\pm
a_1\lambda_2|^2\leq|a_1|^2\left(\lambda_1+\lambda_2\right)^2
=|a_1|^2\left(1+2\lambda_1\lambda_2\right)=|a_1|^2[1+c(\phi)],
\]
and the maximum is attained if $|a_1|^2$ is maximal, i.e.\
$|a_1|^2=1/2$ and $a_2=0$, which completes the proof.

\noindent{\bf Lemma 5.} Let $|\phi\rangle$ be an entangled state and
$|\psi\rangle$ the eigenvector of $[|\phi\rangle\langle\phi|]^{T_B}$ with
the negative eigenvalue i.e.
\begin{equation}
[|\phi\rangle\langle\phi|]^{T_B}|\psi\rangle=-\frac{c(\phi)}{2}|\psi\rangle,
\label{l5a}
\end{equation}
Then $|\phi\rangle$ and
$|\widetilde{\psi}\rangle$ have a common Schmidt basis,
where $|\widetilde{\psi}\rangle$ is the eigenvector of
$[|\psi\rangle\langle\psi|]^{T_B}$ with the negative eigenvalue, i.e.
\[
[|\psi\rangle\langle\psi|]^{T_B}|\widetilde{\psi}\rangle=
-\frac{1}{2}|\widetilde{\psi}\rangle.
\]

\noindent{\bf Proof:} From (\ref{l5a}) we have
\[
-\frac{c(\phi)}{2}=
\langle\psi|~[|\phi\rangle\langle\phi|]^{T_B}~|\psi\rangle
=\text{Tr}\left([|\phi\rangle\langle\phi|]^{T_B}
|\psi\rangle\langle\psi|\right)
=\text{Tr}\left(|\phi\rangle\langle\phi|~
[|\psi\rangle\langle\psi|]^{T_B}\right)=
\langle\phi|~[|\psi\rangle\langle\psi|]^{T_B}~|\phi\rangle.
\]
From Lemma 2 we know that $|\psi\rangle$ is maximally entangled.
Thus, according to Lemma 3, in the last term we can substitute
$[|\psi\rangle\langle\psi|]^{T_B}$ by
$\frac{1}{2}\mathbb{I}-|\widetilde{\psi}\rangle\langle\widetilde{\psi}|$,
consequently:
\[
\langle\phi|\widetilde{\psi}\rangle\langle\widetilde{\psi}|\phi\rangle
=\frac{1}{2}+\frac{c(\phi)}{2},
\]
hence, from Lemma 4, $|\phi\rangle$ and the maximally entangled
$|\widetilde{\psi}\rangle$ have a common Schmidt basis.

\section{}
\noindent{\bf Lemma 6.}
If $\rho$ is an entangled state, i.e. its concurrence
$c(\rho)$ is positive, then
$c^2(\rho)/4$ equals the smallest
eigenvalue of $Y=\Sigma(\rho^{T_B})^*\Sigma\rho^{T_B}$. 

\noindent{\bf Proof:} If $d_1^2/4,\dots,d_4^2/4$ are the eigenvalues of\,\,
 $Y=\Sigma(\rho^{T_B})^*\Sigma\rho^{T_B}$ and $c_1^2\geq\dots\geq c_4^2$ the
(real and positive, see \cite{wootters}) eigenvalues of
$X=\Sigma\rho^*\Sigma\rho$ (c.f. Eq.~(\ref{conc2})), the following relation
holds:
\begin{eqnarray*}
d_1^2&=&(c_1+c_2+c_3-c_4)^2, \\
d_2^2&=&(c_1+c_2-c_3+c_4)^2, \\
d_3^2&=&(c_1-c_2+c_3+c_4)^2, \\
d_4^2&=&(-c_1+c_2+c_3+c_4)^2.
\end{eqnarray*}

Indeed, invoking the anticommutation relations for Pauli matrices, 
we check that
for an arbitrary local transformation $L=U\otimes V$, $U, V\in SU(2)$
\begin{equation}
L^{\ast}=\Sigma L\Sigma.
\end{equation}
We can thus use local transformations to bring $\rho$ in
$X=\Sigma\rho^*\Sigma\rho$ and
$Y=\Sigma(\rho^{T_B})^*\Sigma\rho^{T_B}$ to a relatively simple form. An
arbitrary hermitian $\rho$ can be decomposed as
\begin{equation}
\rho:=\frac{1}{4}{\mathbb{I}}_4+\sum_k\left(a_k^{\prime} \sigma_k\otimes
{\mathbb{I}}_2+
b_k^{\prime}{\mathbb{I}}_2\otimes\sigma_k\right)+\sum_{nm}C_{nm}\sigma_m
\otimes\sigma_n,
\label{rdec}
\end{equation}
with real $a_k^{\prime},b_k^{\prime}$, and $C_{mn}$. By local transformations,
we can bring the $3\times 3$ matrix $C$ to the diagonal form with nonnegative
diagonal elements $\mu_1, \mu_2$, and $\mu_3$ \cite{horodecki2,kus}. The
desired transformation changes $a_k^{\prime}$ and $b_k^{\prime}$ to some other
real $a_k$ and $b_k$, hence finally
\begin{equation}
\rho=\frac{1}{4}+\left[
\begin {array}{cccc}
a_3+b_3+\mu_3 & b_1-ib_2      & a_1-ia_2       & \mu_1-\mu_2 \\
b_1+ib_2      & a_3-b_3-\mu_3 & \mu_1+\mu_2    & a_1-ia_2\\
a_1+ia_2      & \mu_1 +\mu_2  & -a_3+b_3-\mu_3 & b_1-ib_2\\
\mu_1-\mu_2   & a_1+ia_2      & b_1+ib_2       & -a_3-b_3+\mu_3
\end {array}
\right].
\label{rcan}
\end{equation}
Somewhat tedious but straightforward calculations show that
\begin{eqnarray}
&\text{Tr}Y=\text{Tr}X& \\
&\text{Tr}Y^2=\text{Tr}X^2&-\delta_2 \\
&\text{Tr}Y^3=\text{Tr}X^3&-\delta_3 \\
&\text{Tr}Y^4=\text{Tr}X^4&-\delta_4
\end{eqnarray}
where
\begin{eqnarray}
\label{delta2}
\delta_2&=&6d+\frac{3}{2}\text{Tr}X^2-\frac{3}{4}(\text{Tr}X)^2, \\
\label{delta3}
\delta_3&=&\frac{5}{4}\delta_2\text{Tr}X, \\
\label{delta4}
\delta_4&=&\frac{7}{12}\delta_2\left(2\text{Tr}X^2+(\text{Tr}X)^2-
\delta_2\right), \\
\label{d}
d^2&=&\text{det}X.
\end{eqnarray}
On the other hand, as (this time rather short) calculations show, the same
relations hold for two diagonal matrices
\begin{equation}
X^{\prime}=\text{diag}(c_1^2,c_2^2,c_3^2,c_4^2),\quad
Y^{\prime}=\text{diag}(d_1^2,d_2^2,d_3^2,d_4^2)/4
\end{equation}
where
\begin{eqnarray}
\label{dp1}
d_1^2&=&(c_1+c_2+c_3-c_4)^2, \\
\label{dp2}
d_2^2&=&(c_1+c_2-c_3+c_4)^2, \\
\label{dp3}
d_3^2&=&(c_1-c_2+c_3+c_4)^2, \\
\label{dp4}
d_4^2&=&(-c_1+c_2+c_3+c_4)^2,
\end{eqnarray}
if we choose $d=+(\text{det}X)^{1/2}$, or
\begin{eqnarray}
\label{dm1}
d_1^2&=&(-c_1+c_2+c_3-c_4)^2, \\
\label{dm2}
d_2^2&=&(-c_1+c_2-c_3+c_4)^2, \\
\label{dm3}
d_3^2&=&(-c_1-c_2+c_3+c_4)^2, \\
\label{dm4}
d_4^2&=&(c_1+c_2+c_3+c_4)^2,
\end{eqnarray}
if $d=-(\text{det}X)^{1/2}$. Since there is a one to one correspondence between
the set of eigenvalues of a $n$-dimensional matrix and the traces of its
first $n$ powers, the relation between the eigenvalues $c_i^2$ of $X$ and the
eigenvalues $d_i^2/4$ of $Y$ must be given by (\ref{dp1}--\ref{dp4}) or
(\ref{dm1}--\ref{dm4}). The second case, $d<0$, is excluded due to the
positivity of $\rho$. Indeed, one checks that:
\begin{equation}
d=\frac{1}{6}\delta_2-\frac{1}{4}\text{Tr}X^2+\frac{1}{8}(\text{Tr}X)^2
=\text{det}(\rho)\geq 0.
\label{dgeq0}
\end{equation}
The first equality in (\ref{dgeq0}) follows from (\ref{delta2}) whereas the
second is established by an explicit calculation using (\ref{rcan}) and the
definition of $X$ in terms of $\rho$.

Thus the eigenvalues $d_1^2,\dots,d_4^2$ (\ref{dp1}--\ref{dp4}) of $4Y$
are real and positive, and the smallest eigenvalue $d_4^2$ equals $c(\rho)^2$,
see Eq.~(\ref{conc}).   
  
\noindent{\bf Lemma 7.}
If ${\rm rank}(\rho)\geq 3$, where $\rho$ is an entangled state,
the smallest eigenvalue of
$Y=\Sigma(\rho^{T_B})^*\Sigma\rho^{T_B}$ 
is non-degenerate. If $|\phi_4\rangle$ denotes the corresponding
eigenvector, and $|\phi_i\rangle$, $i=1,2,3$, the other three
eigenvectors, the following holds:
\begin{eqnarray}
\langle\phi_4|\rho^{T_B}|\phi_4\rangle & = & -\frac{1}{2}c(\rho)c(\phi_4),\\
\langle\phi_i|\rho^{T_B}|\phi_i\rangle & \geq & 0,\ \ i=1,2,3.
\end{eqnarray}

\noindent{\bf Proof:}
Let $d_1^2/4\geq\dots\geq d_4^2/4$ denote the
(real and positive, see Lemma 6) eigenvalues of $Y$, and
$c_1^2\geq\dots\geq c_4^2$ the eigenvalues
of $X=\Sigma\rho^*\Sigma\rho$. According to Lemma 6, the relation
between $d_i$ and $c_i$ is given by Eqs. (\ref{dp1}-\ref{dp4}),
in particular $d_4=c(\rho)$.
From $c(\rho)>0$ and the definition of concurrence, Eq. (\ref{conc}),
it follows that $c_1>c_2$. Now, if ${\rm rank}(\rho)\geq 3$, it is easy to show
that $c_2>0$ (since ${\rm rank}(\Sigma)=4$ and therefore
${\rm rank}(X)\geq 2$),
and then Eqs. (\ref{dp3},\ref{dp4})
imply $d_4<d_3$. Hence, $d_4^2/4$ is a
non-degenerate eigenvalue. 

By splitting the eigenvalue equation
$Y|\phi_i\rangle=\frac{1}{4}d_i^2|\phi_i\rangle$ (with real $d_i$) into
its real and imaginary part, one can derive that $|\phi_i\rangle$
fulfills $\Sigma\rho^{T_B}|\phi_i\rangle=\frac{1}{2}
e^{i\chi_i}d_i|\phi_i^*\rangle$, where $e^{i\chi_i}$ is a phase factor. 
Using $\Sigma^2=1$, Eq. (\ref{purec}), and the hermiticity
of $\rho^{T_B}$, we
conclude that 
\begin{equation}
\langle\phi_i|\rho^{T_B}|\phi_i\rangle=\pm\frac{1}{2}d_i c(\phi_i).\label{last}
\end{equation}
In order to complete the proof of Lemma 7, it remains to be shown that the
sign on the right hand side must be negative for $i=4$ and
nonnegative for $i=1,2,3$. 
Because of continuity,
it is sufficient to consider the case ${\rm rank}(\rho)=4$.
Then, $|\phi_i\rangle$ cannot be a product vector (since inserting
$|\phi_i\rangle=|e,f\rangle$ into Eq. (\ref{last}) would imply 
$\langle e,f^*|\rho|e,f^*\rangle=0$), i.e. the right hand side of
Eq. (\ref{last}) cannot be zero ($d_i>0$ follows from $d_4=c(\rho)>0$). 
Now, if $\rho$ is infinitesimally
close to an entangled pure state, $\rho\to |\psi\rangle\langle\psi|$, it is
easy to check that, indeed, Eq. (\ref{last}) is valid with the
minus sign for $i=4$ and the plus sign for $i=1,2,3$.
(For $|\psi\rangle=[\lambda_1,0,0,\lambda_2]^T$, one finds that
$|\phi_{1,2}\rangle=[\lambda_2,0,0,\pm \lambda_1]^T$,
$|\phi_3\rangle=[0,1,1,0]^T/\sqrt{2}$, and
$|\phi_4\rangle=[0,1,-1,0]^T/\sqrt{2}$.)
Next, we consider the one parameter family 
$\rho(\lambda')=\mu'|\psi\rangle\langle\psi|+\lambda'\rho_s$,
with $\mu'=1-\lambda'$ and $\lambda'\in[0,\lambda]$, where
$\rho_s$ is the BSA of $\rho=\rho(\lambda)$. Since $\rho_s^{T_B}$ is
positive, $\langle\chi|\rho^{T_B}|\chi\rangle<0$ implies
$\langle\chi|\rho(\lambda')^{T_B}|\chi\rangle<0$, hence
$c(\rho(\lambda'))>0$ for all $\lambda'\in[0,\lambda]$.
Finally, continuity implies that the sign of the 
right hand side of Eq. (\ref{last}) does not change when increasing
$\lambda'$ from $0$ to $\lambda$.

\end{document}